# Randomness and Multi-level Interactions in Biology[1]

Marcello Buiatti

Department of Evolutionary Biology

Florence Uiversity, Florence , Italy

Giuseppe Longo

CNRS – Ecole Normale Sup. et CREA,

Ecole Polytechnique, Paris

http://www.di.ens.fr/users/longo

**Abstract**. The dynamic instability of the living systems and the "superposition" of different forms of randomness are viewed as a component of the contingently increasing organization of life along evolution. We briefly survey how classical and quantum physics define randomness differently. We **then** discuss why this requires, in our view, an enrichment of the understanding of the effects of their concurrent presence in biological system**s'** dynamics. Biological randomness is then presented as an essential component of the heterogeneous determination and intrinsic unpredictability proper to life phenomena, due to the nesting and interaction of many levels of organization. Even increasing organization itself induces growing disorder, by energy dispersal effects of course, but also by variability and differentiation. **Co-operation** between diverse components in networks **implies** at the same time the presence of constraints due to the peculiar forms of (bio-)resonance and (bio-) entanglement we discuss.

## 1. Optimality and predictability.

In his long struggle against creationism and finalism, J. S. Gould tackled the issue of the apparent increasing "complexity" of organisms along Evolution. The question he asked himself was: from prokaryotic to eukaryotic cells and to multi-cellular organisms and within them towards mammals and humans, with many intermediate steps, is there something with one of the many meanings of "complexity" which "increases"? The very well documented answer by Gould is in "Full House", which presents the general argument for denying that progress defines the history of life or even exists as a general trend at all. The theoretical proposal, though largely informal, is entirely based on the role of randomness in the diffusion of life along evolution.

---

[1] ***Submitted for publication.***

The first point to understand within this frame is that, whatever may be increasing, the increase need not to be on a line (a total order), which would give a direction. It may be a partial order, like a tree expanding in all possible directions (but … what does "possible" mean?) or even a pre-order or a pre-ordered network, with crossing-overs between different branches. Different evolutionary paths may give phenotypic analogies or similar paths producing homologies as well as "regressions" in complexity, whatever this notion may mean, with no predetermined orientation.

Exit thus finalism, we also have to drop, in evolution, any notion of "value", or whatever comparison between different organisms which would lead to directionality towards an even local maximum or "best". Our human hand is not "best", nor "better", with respect to any other front podia, from the elephant's to the kangaroo's: they all are just the results of the diverse and possible evolutionary paths followed by the front legs of tetrapods.

If we look back to the history of Biology in the modern era we find that the concept of finalism stems from the birth of the so-called "Modern synthesis" (Huxley, 1943) and particularly in Fisher's mathematical version of it. However, Fisher's dogma (1930) stating that selection works in a stable environment and therefore leads to the evolution of optimal genotypes/phenotypes, has been proven wrong since the pioneering work by Sewall Wright (1932) on the dynamics of populations living in ever-changing environments. Sewall Wright showed that different combinations of alleles can reach the same level of adaptation, introducing the concepts of non additive fitness of genomes and of "adaptive landscape", meaning by that the landscape of fitness of different genotypes. Following Wright then, the concepts of "highest fitness" or the "best fit" have been challenged and we now talk of variable levels of adaptation to variable "environments". That is, adaptation with a co-constituted, far from equilibrium, environment, where individual organisms and the ecosystem are interacting in an ever changing (changing symmetries) and extended, critical transition (our theses, see below).

In the darwinian perspective, it is the less adapted that is excluded. Low or high adaptation are relative and may depend, of course, on the possible presence of another or novel competitive species or variant of/in a population which, locally and on a specific aspect, seems to perform better or worse, in reproduction, possibly in correlation to access to limited resources. However, even minor changes in the environment, below observation at one time, may subsequently show that different causes may render the organism or species more or less adapted, adaptation standing as the only general criteria for natural selection. And in no way, by looking at one specific evolutionary "moment", one may predict which individuals or species will be less or more adapted in … one thousands, or one million years. No way, for instance to predict, at the time, that the double jaw of some vertebrates of the Devonian (the gnathostomes) would yield the bones of the median ear of amphibians (the stegocephalia) and, then, of mammals - a preferred example by Gould. That was a possible, contingent path, by *exaptation* (Gould's original notion), leading to a previously non-existing performance by the new use, after contingent changes, of previously existing organs. Hearing danger and appeals … producing Mozart's music is just contingency, since there was no evolutionary *necessity* to hear nor to write a piano concerto.

Now, which notion of randomness do we need to grasp the biological notions of "contingency" and "possible" (evolutionary/ontogenetic) path?

## 2. Randomness in Physics.

Physicists deal with two forms of randomness, classical and quantum, in separated (actually incompatible) theories, as the classical/relativistic and the quantum fields are not unified: entanglement and Bell inequalities mathematically separate these two theories of randomness (see below).

Too many biologists, since the "Central dogma of molecular genetics" proposed by Francis Crick (1958, 1970) and thoroughly discussed within the frame of a mechanistic theory of life by Monod (1970), think that deterministic means predictable (thus programmable: the theory of the genetic program) and that, similarly as for Laplace, randomness is completely independent from determination[2].

---

[2] Laplace, after contributing to the equational determination of the gravitational systems, by the Newton-Laplace equations, conjectured the complete predictability of "all astronomical facts" (and of generally deterministic systems, see Laplace's work on Celestial Mechanics, from 1799 to 1825). Yet, he was also an immense mathematician of probability theory, as a measure of randomness, which he considered a totally different notion from "determination" (Théorie analytique des probabilités et Philosophie des probabilités, 1812 – 1814). Along these lines, Monod (1970) will oppose "hasard et nécessité", in Biology, whose direct consequence has been the role of the notion of "genetic program", since "computable (or programmable)"

Classical randomness instead has been understood, since Poincaré (1892), as deterministic unpredictability, a consequence of, or mathematically expressed by, non-linear determination. Poincaré started this radical change in the epistemological perspective, by the analysis of the non-linear interactions of just three celestial bodies in their gravitational fields. As a consequence, the Solar system turns out to be non-additive, and the non-linearity of the equations expresses this fact. The addition of one more planet to a single one rotating in a keplerian orbit around the Sun, radically changes the system and its predictability: as intuited, but not formalized, also by Newton, the new gravitational field interferes with the previous one and stability is lost. And "noise" pops out, disturbing the assumed clockwise and predictable dynamics of planets and stars.

More precisely, by his "negative result", as he called it, Poincaré showed that in a simple deterministic system (determined by only 9 equations) stable and unstable trajectories (manifolds) may intersect in an extremely complex way, in points that he calls "homoclines". They intersect infinitely often, but form also "tight meshes ... folded upon themselves without ever intersecting themselves". As early as 1892 thus, Poincaré presents deterministic chaos and its form of complexity for the first time. He then deduced that "prediction becomes impossible . . . and we have random phenomena" (Poincare, 1902), since non-measurable perturbations may lead, because of the presence of homoclines and bifurcations (another notion of Poincaré's), to completely different trajectories (see Longo (2011) for a synthetic presentation and its correlation to computational undecidability – or "non-programmability"). This is why classical randomness is now described as deterministic unpredictability, since "noise", if understood as perturbation/fluctuation below measure, after sufficient time, may measurably modify a perfectly well determined trajectory. Observable randomness in particular appears as soon as a non-linear system has no analytic (i. e. linearly approximated) solutions.

The deterministic chaos thus produced even in our Solar system, allowed the evaluation of the time range of unpredictability for various planets (Laskar 1990; 1994): in rather modest astronomical times, it is unpredictable whether the planets will still be turning around the Sun (from one million years to 100 – as for the Earth). Philosophically, it is fair to say that these phenomena yield an epistemic form of randomness: classically, we cannot access to

---

coincides with "deterministic and predictable", see (Longo, 2011).

phenomena but by approximated measures; even in perfectly determined systems, with time this yields deterministic unpredictability, thus epistemic randomness. By principle, classical (and relativistic, of course) measure is approximated (it is an interval): unpredictability develops along time in view of fluctuations/perturbations below the best possible measure (this entered in Laskar's computations, since he considered the best approximation we can make concerning the barycenters of the planets – a very difficult measure, as they are plastic, both by classical and relativistic principles: no perfectly rigid body may exist).

Can we call the novel unity of at least three planets an "emergent systemic unity"? Is the system complex? Why not? Poincaré describes a very complex dynamics in the phase space. Yet, we shall distinguish, below and very clearly, this minimal form of physical complexity and unity from both the unity and complexity of biological systems, also by their specific form of randomness.

As for quantum randomness, this is objective (or intrinsic to the theory), as it is correlated to quantum indetermination, a very different perspective. Typically, Schrödinger's equation is linear, but it determines the dynamics of a law of probability in a (possibly infinite dimensional) space of functions (the Hilbert space of the state or wave functions), not the dynamics of position-momentum in ordinary space-time. When measure occurs, the equations only predict the probability of obtaining a value. This yields the indeterministic nature of Quantum Mechanics and relates to Heisenberg's indetermination of position/momentum.

Moreover, as a consequence of "quantum entanglement" (Einstein et al., 1935), the very notion of "randomness" radically changes (see (Aspect et al., 1982), (Jaeger, 2009) or (Longo et al., 2010), (Bailly, Longo, 2011) for recent reflections and many references on this). Quantum entanglement was proposed by Schrödinger and mathematically specified by Einstein, Podolski and Rosen in a famous paper of 1935. Einstein's aim was to prove the non completeness of quantum mechanics of view of the incompatibility of entanglement with relativity theory. Entangled quanta are the result of an initial systemic unity (in Schrödinger's state function) of two quanta which are later spatially separated. When their observable properties are measured in remote parts of space, they yield correlated probability values (which violate Bell inequalities, a formal warrant for independent probability), as if instantaneous communication were possible.

In short, if two classical dice interact when flipping (they touch each other or interact by whatever physical way) and later separate, one may *independently* compute the probabilities of their expected outcomes. In contrast to this, there is no way to separate, by measure, entangled particles, even when they are far away from each other, in space. And Einstein's paradoxical deduction, based on a pure mathematical reasoning on Schrödinger's equation, has been shown to be empirically valid: entangled yet remote particles do exist, they have been actually produced many times, since Aspect's early experiments in Paris (Aspect et al., 1982). This is one of the reasons for the incompatibility of relativistic and quantum Physics. There is no internal inconsistency of quantum mechanics as Einstein had hoped (in order to convince the quantum physicist of its incompleteness), just incompatibility with relativity theory. And this is shown by the incompatibility of the two field theories and by this correlation of random phenomena, namely of the observable values, measured in presence of entanglement. This yields different theoretical frames, in physics, as for the understanding of randomness.

## 3. Continua, discrete and quantum randomness in Biology.

Before discussing, in the sections below, about the rich blend of quantum and classical randomness in living organisms, let's briefly analyze some recent use in Biology of the peculiar terms of entanglement and superposition so important in general and particularly for randomness in Quantum Physics. We may actually start by some epistemological considerations based on the history of science in the twentieth century.

As earlier discussed in (Buiatti and Buiatti, 2008), living systems are intrinsically *multiverse* showing at the same time different properties. As an analogy, we will argue that this concept is very near to Schrodinger's superposition. For a long time, starting from the very beginning of the twentieth century, the presence of two contrasting properties both in living and non living matter were interpreted as antinomies and led to harsh debates. In Biology, this was so, in particular, between the supporters of the two aspects, the most intriguing cases of antinomy being "chance and necessity", on one hand, and "continuous and discrete", on the other. At the beginning of the $20^{th}$ century biological discreteness and chance were prevailing, following Mendel's discoveries and the Mutation's Theory by Hugo de Vries, against the holistic darwinian vision of a continuous evolutionary change. In the same years, in Physics, Planck's law and the photoelectric effect and the discrete values of the energy spectrum of the electron seemed to contradict Maxwell's theory of Electromagnetism, based on waves in continuous fields.

The concept of the simultaneous presence of discrete and continuous patterns in Biology was then introduced by the discovery by the Swedish plant geneticist Nillsson-Ehle in 1909 who showed that the continuous distribution of color intensity of wheat kernels was determined by the additive action of variable numbers of color increasing discrete genetic variants (alleles). However, the coexistence of discreteness and continuity as a general feature of living systems was formally discussed and fully accepted later on, by J.L. Lush (1945), K. Mather (1949) and I.M. Lerner (1950). A whole new discipline, Quantitative Genetics, was built on complex mathematical models where also the effect of the environment was taken as a further interfering variable of quantitative polygenic characters.

These findings are conceptually analogous but not homologous to the quantum concept of superposition, both deriving from the existence at the same time in the same object of two contrasting features like discreteness and continuity, but do not materially derive from quantum Physical processes. Similar conclusions can be drawn for some more antinomies in Biology like that already mentioned between chance and necessity or "internal" and "external". For instance, multi-cellular organisms live within ecosystems but have an internal space (Claude Bernard's "milieu intérieur"). One component of it is given by the activities of the lower level of cells. Yet, this internal space can be considered an external one for each individual cell, the molecular network being its internal space, a concept on which we shall discuss more thoroughly later on

As far as entanglement is concerned, this term, defined by Arndt et al. (2009) in a review on Quantum Biology, "means a non-classical correlation", or a non-separability connection between physical observables (or between *physiological properties* as, within Biology, it has been recently advocated by Soto et al., (2009)). Note that, once established, this quantum-like connection may theoretically persist over long distances and times, unless it is perturbed by external interactions and measurements (Arndt et al., 2009). Examples reported by the same authors are polarization-entangled photon-pairs, superconducting circuits, nuclear spins in small molecules, spin noise in atomic ensembles, trapped ions, and other systems (see references in Nielsen and Chuang, 2000).

Quantum Biology is a fairly new but fast evolving inter-disciplinary area of study and obtained very recently a number of interesting data in several biological systems. This should be no wonder, as living systems are made of molecules and quantum chemistry certainly is not a brand new field of research (see for instance the very comprehensive review on proton-coupled electron transfer in Biology by St. Reece and D.G. Noguera (2009)).  For instance, electron tunneling has been observed in cellular respiration (Gray and Winkler, 2003) and electron transport along DNA has been shown by Winkler et al. (2005). The first experimental evidence for proton tunnelling has been given in 1989 (Cha et al., 1989) for the enzyme alcohol de-hydrogenase, which transfers a proton from alcohol to nicotinamide adenine dinucleotide.

Of particular interest are two more areas of investigation with an obviously relevant relationship with the origin of life and evolution, namely quantum related processes in DNA dynamics and in photosynthesis, the major biological process allowing the storage of solar energy and its usage for the construction of living matter. As far as DNA is concerned, Perez et al. (2010) showed that nuclear quantum effects such as tunnelling and zero point motion destabilize rare tautomeric enol forms through the transfer of two hydrogen-bonded protons between adenine and thymine and between cytosine and guanine, thus confirming the suggestion by Watson and Crick in their 1953 paper of a role of prototropic tautomerism of bases in the induction of spontaneous point mutations.  This finding was supported by earlier data obtained in 1995 on the molecular basis of the dynamics and the role of quantum tunnelling in DNA (Doouhal et al.1995). These authors showed through a high resolution analysis of the cooperativity of formation of the tautomer, that the first step of tautomer formation is the shift of protons from one base to the other, a dynamical process of femtoseconds. Morever, M.Noguera et al., 2004, showed that metal cations like Cu 2+ interact with DNA and influence intermolecular Proton Transfer processes. Finally, Ceròn-Carrasco et al, (2009) found that the double proton transfer affects spontaneous mutation in RNA duplexes, particularly in G-C base pairs. This, for biologists, is a particularly relevant discovery because RNA duplexes, according to most theories on the origin of life most probably played a key role in the primeval so called RNA-world.

Photosynthesis captures sunlight energy using the so-called Fenna-Matthews antenna complex containing chromophore proteins present in the chlorosome, the reaction centre. The energy captured is then used it to convert carbon dioxide in biomass. This process has been shown by Enger et al. (2007) to work according to quantum probability laws instead of classical laws. Collini et al. (2010) showed, through two-dimensional photo echo-spectroscopy, quantum coherent sharing of electronic oscillation across proteins at ambient temperature in photosynthetic alga. This result, in contrast with the general idea that the presence of water and high temperatures would determine de-coherence, has been confirmed by Sarovar et al. (2010) who analyzed entanglement in multi-chromophoric light-harvesting complexes and showed that a small amount of long-range and multipartite entanglement can exist even at physiological temperatures.

Recently, quantum biological studies are being extended to other areas as shown for instance by the results of modeling of a possible radical-pair entanglement mechanism of avian magnetic orientation where birds could be "seeing" their path because of the effects of magnetic fields on cryptochrome (see for more details (Rodgers T. et al., 2009) and (Cai J. et al., 2010)).

The general picture coming from all these results confirms the concept that the dynamics of living systems stems from an interaction between classical and quantum processes, the last being possible also at fairly high temperatures and in the presence of water as shown before and confirmed since the early work by Del Giudice (1986). Later, it was also found that stochastic metabolic activities may be induced or accelerated by quantum effects analyzed in terms of Quantum Electro-Dynamics (QED). This may depend on the very peculiar, "(super)-coherent" structure of water in cells, to be also understood in terms of QED (Del Giudice et al., 1998; 2006).

## 4. Biological randomness.

The first point we want to make is that we cannot have a sound theory of biological randomness without at least relating the two forms above of physical randomness: in a cell, the quantum effects may happen jointly to classical dynamics and their proper form of randomness, i. e. they take place simultaneously and affect each other. As far as we know, since "unification (of the fields)" is not yet invented, no physical theory so far deals at once with the "superposition" of classical and quantum randomness.

Thus, beginning with intracellular phenomena and then with cell to cell interactions, in a tissue, one may need a classical, a quantum but also an ad hoc dynamical treatment, with largely stochastic dynamics whose structure of determination (and subsequent randomness) is far from being known. And, this is not the end.

### 4.1 Increasing organization.

In living systems, the formation of different levels of organization, both in embryogenesis and evolution (see later), corresponds, in principle, to a *local* decrease of entropy, since organization increases, thus leading to a decrease of randomness. That is, during these biological processes, locally, some more "order" appears in the Universe – at the price, of course, of energy consumption and, thus, of increasing entropy somewhere else. Yet, each of these "new" levels of organization has its-own non-linear (or quantum) internal dynamics and, moreover, (and this will be crucial for our analysis) they mutually interact. The novel forms of interactions, made possible by the differentiation, yield a new form of randomness, which we will understand in terms of "bio-resonance" in sub-section 4.2.

In order to discuss these two contrasting and coexisting tendencies of life – formation of order while inducing new forms of randomness – we will first discuss the processes of increase in number, differentiation and connection of components, throughout evolution from the possible origin of life to the formation of the interacting and interwoven hierarchical organisation of life in the Biosphere.

As discussed thoroughly in a review by Buiatti and Buiatti (2008), the global living system is hierarchically organised into levels of increasing complexity, all with some general properties but also with level dependent ones (see also Bailly and Longo, 2011). Most probably, everything started when two kinds of macromolecules, DNA and RNA, built a dynamical micro-system capable of reciprocal replication as described by Eigen and Schuster (1979). The next step may have been compartmentalization and the formation of primeval cells. Then, aggregation processes gave rise to *colonies* whose cellular components were partially endowed with a nucleus or anyway with what we call now an eukaryotic organisation of the genome. Later on, fusions between nucleated cells and others bearing two kinds of small circular genomes, the mitochondria, capable of respiration, present in animal cells, and, in the case of plants, chloroplasts, capable of the only process present in living systems leading to solar energy fixation

Multi-cellular organisms belonging to both plants and animals kingdoms were later born and a cooperative "division of labour" was developed in single organisms between groups of cells capable of different functions, through a complex process of multiplication and differentiation. This hypothesis is supported by a number of examples in colonies of bacteria and of eukaryotes, along developmental lines which resemble cells and tissue differentiation in multicellular organisms, all showing composite unicellular/multicellular life cycles.

Already in 1979, J.W.T. Winpenny and J.A. Parr observed different levels of activity of a series of enzymes in the external ad internal areas of *Enterobacter cloacae* large colonies. Later on J.C. McMichael (1992), studying the behaviour of *Moraxella bovis* colonies grown between agar and polystyrene in Petri dishes found that they were differentiated into concentric rings The two outer ring zones yielded dividing bacteria that formed agar surface colonies of very different morphology from that of the innermost zones, where cells were quiescent. Rieger et al., 2008 and Cepl et al., 2010, studied thoroughly the morphology and behaviour of *Serratia marcescens* "bodies" as they called them, with an explicit reference to multi-cellular organisms morphology. Also *Serratia rubidaea* bodies were found to be differentiated in areas of different colours and textures and differentiation was found to be directed by a series of molecular signals between cells and coming from the outer area of the colonies. Differentiation then was shown to be induced by a *dynamic cooperation* between cells, where varying concentrations of signals in different areas of the colonies regulate both texture and colours.

It is worth noting that, as we shall discuss later, the dynamic flow of molecular signals also induces the primary differentiation in the embryos of most multi-cellular animals from *Drosophyla*, where it has been fully demonstrated for the first time, to mammals. In all these cases, thus, differentiation is mainly epigenetic, meaning by that that it is generally not due to changes in the genetic complements of the aggregating cells but to different levels of expression, from zero to a maximum, of different sets of genes in different areas of the "organism", where by organism we now mean all kinds of organisation of living systems from cell colonies to plants and animals. This possible model of the intermediate steps from unicellular to multi-cellular organisms seems to be supported by two examples, a prokaryotic and an eukaryotic one, in whose life cycle, "bodies" endowed with complex cell differentiation derive from the aggregation of single cells (see the review by Dee et al. (2000)).

The two organisms are *Myxococcus Xanthus,* a prokaryote, and *Dictyostelium discoideum,* an eukaryote (an amoeba) both living in the soil and feeding on bacteria. Both have surprisingly similar general life cycles involving a process of aggregation, the use of sensor histidine kinases to regulate development, and the use of mechanisms of quorum sensing to count the number of cells before multicellular development. Bacteria and amoebas live for a large part of their cycles as individual cells, yet, while Dyctiostelium feeds by phagocytosis, Myxococcus attacks other bacteria in groups using large quantities of digestive enzymes. Therefore, while in some bacterial cycles cooperation between different cells occurs also before aggregation, this collaborative behaviour in Dictyostelium develops later. Aggregation is induced when nutrients start being scarce and cells tend to join and aggregate with a "founder" one through complex systems of chemiotaxis. The process is induced and regulated by the induction of cyclic AMP synthesis by low concentrations of food. The "founder" amoeba will then be the first of the population whose internal level of cyclic AMP reaches a threshold level. At this moment AMP production enters into an exponential phase and a part of it is released in the medium inducing the movement of surrounding amoebas, which will aggregate forming initially undifferentiated mounds containing an average of 100,000 cells. The mound will then differentiate leading to the formation of a slug which will move in search of a new, more favourable environment in terms of amounts of nutrients. Once found such a suitable environment, the slug will stop its search and transform itself into a stalk and a fruiting body where cells will be changed into spores to be released. Each spore therefore will become again an amoeba and a new cycle will start. The cycle of Myxococcus is similar although the fruiting structures are much less complex.

Now, it is worth noting that the number of cells in the slug of Dictyostelium is far higher than those which will become spores giving rise to the amoebas of the future cycle, and this suggests the possible presence of selection processes. As a matter of fact, amoebas single cell populations are highly heterogeneous, a fact which prompted a series of studies of the dynamics of what may be called the "social structure and dynamics of slime moulds". Now we know (Ostrowsky et al., 2008, Flowers et al., 2010, Mehdiabadi et al., 2008, Strassmann et al., 2000) that the aggregation processes are based on kin recognition and on the so-called "cheaters", that is genetically different cells generally selected against and therefore entering only in a small number in the fruiting body and in the next generation. As shown for instance by N.Van Driessche et al, (2002), about 25% of Dictyostelium genes are differentially regulated during development and in different cell types through a cascade of signal transduction events initially triggered by the variation in cAMP concentration in time and space, based on coordinated cell to cell interactions. Differentiation "downwards" processes therefore are very similar in Dictyostelium to those occurring in all animals and plants, all depending on initial triggers coming from cell environment and from epigenetic cell-to cell interactions. This pattern is very similar to what happens again at the population level in other kinds of social systems such as bees', for instance, or other societies of insects. In this case physical changes and behavioural dynamics are both epigenetically induced by interactions between individuals belonging to the same community. In the case of bees, typically, cast differentiation has been shown to derive from social behaviour as shown by the fact that "queens" and workers have the same genotypes, while queens show strikingly different gene expression patterns derived from the nutrition by workers with the so-called royal jelly (see for instance Kucharski et al., 2008).

Climbing to yet higher order systems, we are all aware that further levels of organisation between different mono-specific populations of organisms (the "avatars" of Eldredge (2008)), lead to the formation of ecosystems all connected to each-other in the same continent and between continents trough air and water.

We next discuss in which sense increasing organization yields further forms of randomness. In section 7, we will try to understand the role of random evolutionary paths as a possible understanding of the increasing complexity of life, including the unicellular aggregations discussed above.

**4.2 Bio-resonance.**

A cell is a network of interacting molecules in a membrane, colonies and tissues are made of groups of interacting cells, individual organisms are interacting with others of the same and different species in the same ecosystem and all ecosystems interact in the Biosphere. Interactions thus happen at all levels of organisation between different objects within that level, and, at each level, they are not additive. However, as we shall see later, communication is both intra and inter-level and therefore a change in one level may spread its effects at the same time to higher ad lower ones. At the molecular level interactions between molecules generally lead to the formation of complexes between two or more molecules with globally different and complementary conformations (as for their non additivity, see (Ricard, 2008)), while cells interact either exchanging molecules through intercellular membranes or through the reciprocal perception, as it happens for instance in the case of contact inhibition between differentiated cells. Contact inhibition is critical for the maintenance of the organisation of cell networks particularly in animals where disruption may mean undesirable cell division putatively leading to cancer (Soto and Sonneschein, 1999). Thus, exchange of molecules may be "horizontal" (within the same organisational level), "vertical" (between levels) or between the organism and the external environment, and it is mediated by the recognition of specific molecules or energetic inputs by complementary trans-membrane receptors. After recognition the receptor will change conformation and start a cascade of molecular transfers within the cell, very often leading to gene activation or repression and the active synthesis of new molecules needed for the response to the primary incoming input.

We have, therefore, in logical terms, a "higher order" form of intra- and inter-level interactions, rarely dealt with (if ever) in mathematical Physics. Poincaré's work allowed to focus on *resonance effects* between two planets and one Sun: just one (relatively simple) deterministic system or, one simple level of organization with non-linear interactions. And he invented deterministic chaos[3]. Quantum Mechanics deals with quantum randomness at its own scale, as we observed, and unification with relativistic/classical physics is far from accomplished. In short, a lot may be said in physics even when looking at only one level of determination/organization.

On the grounds of the phenomena discussed above, in Biology instead, one has to deal with interacting, yet different levels of organization, each possessing its own (possibly mathematical) form of determination, which may be non-linear or quantum. This multi-level interaction induces what we may call "bio-resonance" and its subsequent proper form of randomness. Can this be reduced to a familiar form of physical randomness? Why not: we let the job to the reductionist, who should first unify, though, or at least correlate quantum and classical randomness. For the time being, we point out the physical singularity of life and this peculiar phenomenality of randomness as a consequence of higher order interactions and we try to conceptualize it as clearly as we can. This taking into account that even computational and dynamic randomness have been only recently (and partially) correlated, see (Longo, 2011).

In short, bio-resonance takes place between different levels of organization, each level having its "autonomous" activity liable to be treated with different conceptual - sometimes mathematical – tools, each with its own form of (internal) non-linearity, resonance and alike. Moreover, each system or level is integrated and regulated by all the other elements of the

---

*3* Technically, in Astrophysics, two planets are in maximal resonance/gravitational interference when they are on the same line with respect to the Sun. Many other forms of resonance, as a component of "divergence" of possible trajectories and, thus, of unpredictability, have been analyzed in mathematical physics. A rather general one is the Pollicott-Ruelle resonance, which applies also to open systems and is related to various forms of dynamical entropy, (Gaspard, 2007). Thus, while the first form of instability (Poincaré's) is analyzed in system at equilibrium, the second form may be extended to systems far from equilibrium.

*same* as well as of *different* levels of organization - while affecting them. The other fundamental aspect we want to stress here is that the bio-resonance, based on integrating and regulating interactions, has both a stabilizing and a destabilizing role, in contrast to the physical "parallel" notions, mostly related to divergence and instability (both in astrophysics and in more general (thermo-)dynamical systems, see the footnote above).

For a better understanding of the basics of this discussion we shall introduce here some concepts typical of living systems, analogous but not necessarily homologous to similar ones which are well-known in structures and dynamics of non living matter. Biological objects are, as discussed by Waddington, "homeo-rhetic", as opposed to homeo-static, in the sense that they fall back into a non-living state if their components are not able to maintain a good level of dynamical connection within levels of organization.

In view of the randomness proper to each level and of the one due to bio-resonance between different levels, they move along unpredictable paths but within the limits allowed by the structure and dynamics of the connections (including regulation and integration, both being bio-resonance functions). According to Arndt et al. (2009), these connections between components and levels of organization can in a way be also analyzed in terms of the concept of entanglement or, may be better, "bio-entanglement", meaning by that non microphysical effects. In a similar way, we could speak above of bio-resonance when the stabilizing/destabilizing role yields coherence not within but between different levels of organization. In all cases, it should always be remembered that the specific and general structural and dynamical "rules" of bio-entanglement and bio-resonance have been and are evolving in time under the pressure of processes of "extended selection", a term recently suggested by one of us (M. Buiatti, 2011, in the press) including classical selection by the environment in a broad sense, but also "internal selection" (G. Bateson, 1979), that is selection discarding changes not complying coherence rules within the organism. Moreover, each living system is endowed with robustness and resilience whose structure is continuously changing and that can be considered in an ongoing critical transition - the notion of "extended criticality" in (Bailly and Longo, 2008; Longo and Montevil, 2011): a cell, an organism … is always poised on a critical threshold.

This is why, the analysis of biological processes in terms of interactivity and non-linearity is sound, but still insufficient as both in classical and quantum dynamics there surely are plenty of interactive, non-additive systems, all major challenges for mathematical intelligibility, yet their complexity is not comparable with the biological complexity (in terms of critical transitions, any reasonable measure of the latter is "infinite" in mathematical terms, see (Bailly and Longo, 2011)). For instance, the networks, as analyzed by Barabasi (2004), are largely inspired by computers' networks, with their hubs and "scale invariants". Their analysis is often limited to the definition of structures from a formal point of view, without entering into the discussion of the dynamics of living systems. In particular, they only deal with just one level of organization, determined by the mathematics of networks, based on two key assumptions: a principle of "preferential attachment" and the maintenance of the "mean number of connections" at each node, see (Fox Keller, 2005). It is then possible that the world-wide web and the cellular metabolic networks, when abstracted form their multilevel contexts, yield similar results and are shown to obey to the same rules. Even when dynamical functions are added to the "small-world" models, simulation experiments yield results common to "classical" physical non-linear system. This may be very informative for understanding some very general behaviors, much more than, say, recognizing that a falling dog is accelerated like a stone (due to another major physical invariant), yet it similarly abstracts only a specific physical dynamic. The situation is similar as for critical phenomena, where universality laws do give relevant information on common feature of critical transitions (Bak et al., 1988). The same should be said for the continuous dynamics of morphogenesis that, since Turing (1952), and by completely different mathematical tools, opened the way to very important analysis of processes of development and organs formation, see (Jean, 1994). Yet, these approaches, though very interesting, are based on strong physico-mathematical assumptions and deal with just one mathematical structure of determination. Thus, they stop on the verge of the proper complexity of the heterogeneous, many level organization of living entities.

As a matter of fact, the biological challenge begins when one deals with a metabolic network in a cell, which belongs to a cellular network composing a tissue, which is part of an organ, integrated in an organism and so on, as discussed before, where all the different levels of organization are inter-regulated by cascades of events in all directions, exposed and reacting to matter and energy inputs coming also from the non-living matter components of the systems. Moreover, all changes involve interference and superposition between classical non-linear and quantum processes.

This is the conceptual, first, then the mathematical analysis, if possible, we have to work at, in order to grasp the proper complexity of the living state of matter, well beyond, but including, the non-linear, interactive structures of physical dynamics and networks.

**5. Complicated/complex**

A non-linear system may be also obtained by a bottom-up artificial construction: consider a double pendulum, connected to a spring pulling on a non-linear oscillator ... this yields a highly unpredictable non-linear machine. And one may also compose many of these devices into a network of such machines … and obtain even more complicated structures. Is such a system complex or just "complicated"? Yes, the machine is complex, if non-linearity of interactions is considered a sufficient characterization of complexity. No, it should be better called "complicated", if we consider *complex* an object only when it has a many level structure of determination, each level being understood by different conceptual/ mathematical tools, and when it is, at once, "ordered *and* disordered, regular *and* irregular, variant *and* invariant, stable *and* instable, integrated *and* differentiated" (an understanding of complexity proposed by Edelman). We shall add to this list of conceptual oppositions, "far from equilibrium *and* maintaining some components at equilibrium, dissipative *and* conservative, auto-organized *and* subject to constraints, entropic *and* anti-entropic, in critical singularities *and* extended to an interval of criticality" (see below). Some networks, say, may satisfy part of the first list of opposite properties, yet physical networks are treated as equilibrium systems and do not satisfy the second list of conceptual oppositions.

As a matter of fact, the latter co-existing opposites may be properly found only in presence of different levels of organization. By this we mean, molecular, cellular, tissues' levels, composing organs of an organism and so on, where integration and regulation are due to the key fact that all levels with the exception of the molecular level are constructed top-down**,** their "components" being a priori integrated as they originate from a pre-existing organization. If we take as an example the organismal level, integration and regulation are not due to addition, but to differentiation from an original "organism", one zygote cell. No way to construct a multi-cellular organism by adding a leg or a brain, as we can add a wheel or a computer to the pendulum/spring/oscillator device above. Or, as we can add further nodes, hubs and links to an artificial network, the W.W.Web for example. Note that in the case of the slug of Dictyostelium mentioned in sect. 4.1, there may be an extension of the "organism" by new cells, even non-kin ones (the "cheaters"). Yet, most of the genetically non homogeneous ones will be excluded and their addition does not yield the addition of an "organ", but at most of individual cells which will be eventually transformed into (part of) an organ, only after differentiation. The constraints due to the global structure contribute to induce differentiation, thus (and because of) variability. Or, as always in biology, constraints induce variability *and* variability induce constraints. Thus, even in the formation of colonies, as organismal structures, differentiation is essentially an epigenetic phenomenon which applies to one or a few genetically kin cells (except for a few cheaters).

It may be thus fair to call "complicated" the artificial devices liable to be constructed essentially bottom-up, and reserve the word "complex" to living structures, in particular because of the phenomena hinted above:

1. the top-down construction, as differentiation and integration from one or a few (genetically homogeneous) individual cells;
2. the bio-entanglement;
3. the bio-resonance.

The last two properties are made possible by the first one, which forces integration, while bio-entanglement and resonance are also part of the constraints and, thus, contribute to differentiation.

As for one more instance of bio-entanglement, consider that organs, typically, in an organism are "physiologically entangled" with each other and with the organism: physical entanglement is expressed in terms of *quantum measure*, similarly, there is no *physiological measure* of an organ's activity which may completely pass by the superposition with the others' activity. No reasonable physiological measure can be made on a brain isolated in a flowerpot. Does a lung or a vascular system make sense away from an organism and the combination of its physiological activities?

Nesting and coupling different levels of organization, proper randomness (due to classical/ quantum coexistence and to inter-level bio-resonance effects), entanglement of components as superposition of physiological activities …. The blend of all these peculiar conditions yields the proper complexity of the living state of matter.

**6. More on biological randomness.**

As mentioned above, classical randomness is deterministic unpredictability in equilibrium dynamics. In case of non-linear interactions, the system may yield highly unstable trajectories, yet constantly remaining at equilibrium. These systems are conservative and their trajectories may be computed in principle (yet not predicted) by extremizing a functional (a lagrangian or hamiltonian), that is they go along an optimal trajectory in a suitable phase space, a geodesic. In an equilibrium dynamics, a trajectory is stable when it is not modified by minor fluctuations/perturbations. It is unstable, when it is subject to the too well known (and so badly understood) "butterfly effect", whose mathematics we owe first to Poincaré and his analysis of the planetary determination. This effect is more rigorously called "sensitivity to border conditions", but two more mathematical properties are required as for classical deterministic chaos, as better specified in the 1970s, (see Devaney, 1989).

In Biology, it is not a matter of stable or unstable equilibrium, but of *far from equilibrium*, yet "structurally stable" systems, two very different notions. This hard to understand simultaneous structural stability and non conservative behavior is a blend of stability and instability due exactly to the coexistence of opposite properties such as "order/disorder … integration/differentiation" mentioned above. Typically, integration and differentiation stabilize and destabilize, preserve and modify symmetries, since they maintain the global stability, while allowing the never ending reconstruction of the living system as exemplified by the organism, by continually passing through "critical states". An organism is not "just" a process, a dynamics: it is a permanent passage through a critical state; it is critical in the sense, in particular, that it continually changes symmetries by breaking existing symmetries and constructing new ones (Longo and Montévil, 2011).

As a matter of fact, each mitosis breaks symmetries: the two "new" cells differ from the originating one and are not identical. Mathematically, this changes the invariants, yet it may preserve the ever changing homeorhetic stability of a global structure, be it an organism or an ecosystem, while exploring new paths, by differentiation and variability. This exploration of possibilities takes place, along evolution and ontogenesis, in unicellular organisms as well as in embryogenesis, in the permanent reconstruction of a multi-cellular one, as well as of a higher order system.

But … which "possibilities"? Let's focus on evolution as this is where the novel challenge of biological randomness more clearly steps in. In contrast to all known forms of physical randomness, including quantum randomness, the phase space of possibilities is not "already given", but rather, it is randomly co-constituted, a major mathematical challenge. By the superposition of classical and quantum randomness, in a cell, an organism, an ecosystem, and by bio-resonance of different levels of organization, we need, in Biology, a third, stronger, notion of randomness. A form of randomness where the very list of possibilities to be explored is not already given as it usually is in Physics. When flipping a coin or throwing dies, we know in advance the list of two or six possible outcomes. Even in quantum Physics, the list of possibilities is mathematically pre-given, be it an infinite space like a Focks' space that accommodates all the highly unpredictable, but possible creation of new particles. In Biology, it is the very "phase space" of evolution which is randomly co-constituted – and this mathematical challenge is not treated by current physical theories.

Moreover, in reference to Physics, many are still talking of "equilibrium dynamics" (and we also did so in reference to deterministic chaos) and molecular biologists are hardly digesting Poincaré's discovery that an equilibrium dynamics may be highly unstable, thus random, yet deterministic. As we said, the still dominating Monod's alternative (Nécessité et Hasard), suggests that deterministic means "necessary" thus predictable (i. e. programmable) and that randomness is a very different issue, due to (mostly external) noise.

From our perspective, in Biology, the situation is rather different. As we recalled above and it is widely acknowledged (but not always applied), living objects yield always far form equilibrium, dissipative systems. This is so for one single organism as well as for an ecosystem. In some cases the system may be stationary or in a steady state (constant input/output flow of energy), or, at least, this is a reasonable working assumption, in order at least to write thermo-dynamical balance equations. In this mathematical assumption, some aspects of evolution were analyzed (Bailly and Longo, 2009) and the notion of "extended critical transition" was proposed (Bailly and Longo, 2008; Longo and Montevil, 2011). In this context, the dissipative, far from equilibrium structuring of living matter, may propagate downwards, to the molecular level or upwards to the higher order levels: that is, non-equilibria may downward or upward affect and destabilize or, more soundly, des-equilibrate other levels. For instance in the case of interacting cells (in a tissue, say), "disorder" (or the failure of the cells' dialogue) can propagate downwards to the physical dynamics of molecules – or it may even derive from the organism (under the form of stress, say). Conversely, as everybody agrees, disorder may go from molecules to the whole organism, the latter being the key component for the exploration of diversity, proper for instance to evo-devo systems. Let's consider a few examples of upward/downward bio-resonance inducing an increase in variability.

As discussed at length by Zeh et al. (2009), eukaryotic genomes contrary to prokaryotic ones have a very low relative amount of sequences coding for RNA or proteins: non-coding ones cover the vast majority of DNA. In humans for instance coding DNA is only around 3%, the rest being non-coding but containing a large number of transposons, that is mobile elements capable of "jumping" from one location to another in the genome, inserting themselves and dispersing their copies in a quasi-random distribution. A condition for the insertion is the production of cuts in the receiving sequence thus inducing changes in it. Moreover, some classes of transposons, such as the so-called "helitrons", often carry with them parts of the sequence of the original insertion site re-arranging the genome and changing the order of genes. This often means permanent changes in expression levels of the host sequence with putatively relevant heritable changes in the phenotype of the organism affected. Now, transposons are generally kept silent by the organism through a process involving the addition of a methyl group to their DNA, but in the case of stress de-methylation occurs and transposon mobility is restored suddenly increasing mutation frequencies in a quasi-random way because the only constraint to randomness is given by the fact that in general transposons tend to insert themselves in GC-rich regions of DNA.

The result of this process is therefore the production of new variability which may affect the functionality of genes endowed of critical functions as, for instance the tumor suppression genes whose inactivation may lead to cancer. In this case therefore inter-level resonance is both upward and downward. External stress from the environment induces transposon jumping within cells (downward) but the induction of randomness, through transposon activation in cells, will change the cell itself and it may induce its proliferation and therefore cancer with higher frequency. These events happen in cells but they will have negative effects on the whole organism as a consequence of this inter-cellullar, intra-organismal bio-resonance but are induced by an external input (inter-level bio-resonance).

Another interesting example of bio-resonance may be found in the well known process of development in animals from the fertilized egg to the whole organism (for a thorough discussion of animal development, see Gerhart and Kirscnher (1999). In Drosophyla, but also in mammals like us, the developmental process is activated by the "injection" into the fertilized egg of RNAs and proteins codified by maternal genes and coming from mother cells in an antero-posterior and dorso-ventral direction, thus generating concurrent concentration gradients leading to an oscillator. This induces the formation of sects within which other gradients are formed regulating the differentiation of cells all being submitted, along the two axes, to different ratios between the regulatory molecules. In this case the inducing molecules come from the upper level (the body of the mother), and activate the lower cellular level inducing differentiation. This is surprisingly similar to the phenomenon discussed above concerning Dictyostelium, where gradients induce differentiation and allow to establish new connections, thus new organization, as integration of differentiated cells. On the other hand a "pure" upward effects of bio-resonance can happen when a mistake in DNA replication may induce a mutation in a differentiated cell leading to cell division and inducing cancer, a syndrome dramatically modifying the whole organism and its interactions with the environment. The opposite case may be a downward process like a change in a few connections between neurons, a process not happening only as a response to external agents but deriving from the internal dynamics of the brain. A change like this, of course, being the brain the central controller of the whole body, may modify several processes in an unpredictable number of body components.

Of course, these interactions are highly random not only because of the non-linear effects between the molecules injected and the cells (and of the possible quantum indeterminism discussed above), but also because the whole phenomenon is submitted to environmental changes, which cannot be described, let alone predicted, by focusing only on any of the other levels of determination and randomness, that is on molecular cascades or on cells or even on developing organism as a whole. In conclusion, this many-folded form of randomness in Biology is at the core of variability, diversity, thus of evolution and development.

**7. Anti-entropy**

In Biology, there exist two forms of production of entropy, as disorder. One is due to the many thermodynamical processes that occur at all levels of organization where there is an energy flow and/or production. By the second principle of thermodynamics, applied to far from equilibrium, open (dissipative) systems, these flows and activities produce energy dispersion, thus physical entropy. The other form is properly biological and is due to *variability*, both as individual and evolutionary variability and as differentiation (these two notions are related, of course). This second form of "increasing disorder" may coexist with increasing order: when one has two cells instead of one by mitosis, the order, locally, increases. Yet, since, along reproduction, cells always variate and/or differentiate, this process induces *also* some disorder (beginning with random differences in the proteomes' composition, see (Longo, Montévil, 2012)). Of course, the reductionist may tell us that this fact may be understood in physical terms. Yes, but no physical theory, so far, deals with non-identical reproduction as a core theoretical invariant. Crystallography, say, analyzes crystals' generation and their symmetries, but "imperfection" is hardly the main theoretical invariant. In Biology, instead, both variability and organization need to be theoretically described, in their multi-level structuring, in order to understand the survival and evolution of organisms (Buiatti, Buiatti, 2008). In a synthetic way: biology is largely the *iteration* of a *never identical* morphogenetic process.

In order to deal with these opposing concepts, Bailly and Longo (2009) have introduced the notion of *anti-entropy*. This is a proper observable of life, quali-quantitavely represented by the number-theoretic evaluation of the complexity of

- cellular networks (number of nodes, hubs, links …);
- the fractal dimension and number of connected components of organs[4];
- the number of tissue's differentiations

(we refer to the paper for a more formal treatment of these quantities).

This is a largely incomplete list of items of what could be called proper *biological complexity* and provides just a preliminary "mathematical skeleton" of the much richer complexity of life. One may add lots of further phenotypic properties that popped out along evolution: networks within and of groups and populations (ants or bees' organizations, say) and much more. Its increase or just its maintenance, corresponds to a decrease of entropy (anti-entropy increases or maintains organization, by opposing entropy increase). Yet, anti-entropy is not negentropy: typically, the sum of an equal amount of entropy and negentropy gives 0, a simple singularity, in all the theories where the latter is considered (Shannon, Brillouin, Kolmogorov …). One finds positive anti-entropy only in a living entity, where it adds to the inevitable production of entropy in a non-obvious singularity, that is in extended critical transition in Bailly and Longo, (2008), Longo and Montevil (2011). The name, but just the name, is inspired by anti-matter, as the sum of matter and anti-matter never gives 0,

---

[4] To give a few examples, mammals lungs have a fractal dimension greater than 2, while the lungs of a frog have a 2-dimensional surface. As for organs' "connected components", primates may have up to 600 muscles, while a horse or a cow at most 400.

but the production of gamma-ray with the double of the energy of the annihilated particle.

As far as evolution is concerned, Bailly and Longo (2009) transferred Schrödinger's analysis of quantum dynamics from the dynamics of a "law of probability" to an analysis of phenotypic increasing complexity along evolution as the dynamics of a "potential of variability". Schrödinger's equation may be actually understood as a diffusion equation of the state function over a very abstract space (a Hilbert space – far away from physical space). The idea is to describe the increase of biological complexity as the random diffusion of the density of biomass over time and anti-entropy, yet another very abstract space phase (diffusions, in physics, usually takes place in time and *space*, or of matter within another matter – thus in space).

By quali-quantitatively describing biological (phenotypic) complexity as anti-entropy, it has been possible to reconstruct Gould's phenomenal curb of increasing complexity in evolution as the random diffusion of increasing biomass over time and anti-entropy, a way of specifying this dynamics of a potential of variability, restricted to the diffusion of bio-mass over this peculiar space of observables. This mathematizes Gould's intuition that the random diffusion of life entails increasing complexity just because of the original symmetry breaking due to the formation of the first living entity, which is considered by principle, of "least complexity" (an arbitrary, but sound, axiomatic choice). Gould calls this asymmetry, the "left wall of life", its origin, whatever this may have been. In general, any diffusion is based on random paths which randomly propagate, by local interactions, the initial symmetry breaking, due, in this case, to the existence of the "left wall". Humans, thus, are just one of the possible outcomes of the random complexification of bacteria, via many intermediate random explorations, mostly unsuccessful. Yet, these random explorations are, on average, slightly biased towards increasing complexity, as they propagate the original asymmetry. That is, occasionally, the random distribution of changes yields a more complex structure, slightly more often than simpler one**s**. Intuitively, the ecological niches of simpler organisms are slightly more occupied, thus another "simple" new organism has slightly less chances to survive than a more complex one. And this is so just by local "bumps" towards the right, similarly to the local asymmetric bouncing towards the right of the particles of a gas diffusing from an explosion against a wall on the left.

Gould (1998) has no sound representation of time in his empirical drawings of the "biomass over complexity" curb. By a dual use of time and energy with respect to Schrödinger's operational approach, Bailly and Longo (2009) could explicitly give the time dependence of the curb (in short, time is considered as an operator and energy a parameter, the opposite of what is done in quantum mechanics). And the curb fits the original explosion of bacteria, as well as the fact that bacteria remain the relatively dominating biomass, still now.

Of course, as we said, a random diffusion means that complexity may equally decrease, yet, on average, over time, it does increase. It also means that there is no aim, but pure contingency, and that this contingency is local, by a random, but slightly biased local effect – no global orientation towards increasing complexity, no aim, no target.

In conclusion and in order to go back to the main theme of this paper, one may soundly understand the complexification phenomena described above in terms of random processes. For example, the formation of organism-like colonies of differentiated and integrated bacteria or unicellular organisms, is the result of the *random exploration by variability* of a possible structuring of life. Most explorations fail; those which yield an (even slightly) more complex structure, have (slightly) more chances to be viable, by the propagation of the diffusive asymmetry explained above.

But which mathematical form of randomness are we talking about? It does not matter, as for this global, largely qualitative analysis of evolutionary complexification. Yet, a more precise theoretical or even mathematical specification of this novel form of randomness, possibly along the conceptualization hinted in the previous sections, in ters of bio-entanglement and bio-resonance for example, would greatly help us to better understand the dynamics of life.

-----------------

**References** (Longo's papers are downloadable from:  *http://www.di.ens.fr/users/longo/* )

Arndt, M., JuffmannTh., Vedral, V., 2009. Quantum Physics meets Biology, HFSP Journal, Vol. 3, 6, 386–400,

Aspect A., Grangier P., 1982 Roger G., Experimental Realization of the Einstein-Podolsky-Rosen-Bohm Gedankenexperiment : A New Violation of Bell's Inequalities, Phys. Rev. Let. 49, p.91

Bailly F, Longo, G., 2007, Randomness and Determination in the interplay between the Continuum and the Discrete, Special issue: Mathematical Structures in Computer Science 17(2), 289-307.

Bailly F., Longo G., 2008 "Extended Critical Situations", in J. of Biological Systems, Vol. 16, No. 2: 309-336.

Bailly F., Longo G., 2009, Biological Organization and Anti-Entropy, J. Biological Systems, Vol. 17, No. 1, pp. 63-96.

Bak, P., Tang, C., Wiesenfeld, K., 1988, "Self-organized criticality", Physical Review A 38: 364–374.

Barabasi A.L., Oltvai,Z.N., 2004, Network Biology: understanding the cell functional organization, Nature Reviews Genetics:101-11.

Bateson G., 1979, Mind and Nature: A Necessary Unity, Bantam Books.

Bell J.S., 1964, On the Einstein-Podolsky-Rosen Paradox, Physics, 1, p.195-200

Buiatti, M.,2011, Plants: individuals or epigenetic cell populations?, In: The future of Lamarckism, M.I.T., Press, in press

Buiatti M., M.Buiatti, 2008, Chance vs. necessity in living systems, a false antinomy, Biology Forum,101, 29-66

Cai, J.,Guerreschi, G.G., Briegel, H.J., 2010, Quantum control and entanglement in a chemical compass arXIV:0906.2383v4 (quant-ph).

Čepl,J.J., Patkova, I., Blahůškov,A., Cvrčkov,F., Markoš,A.,2010, Patterning of mutually interacting bacterial bodies: close contacts and airborne signals, BMC Microbiology, 10,139.

Ceron-Carrasco, J.P., Requena, A., Perpete, E.A., Michaux,C., Jacquemin,D., 2009, Double proton transfer mechanism in the adenine-uracil base pair and spontaneous mutation in RNA duplex., Chemical Physics Letters, 484, 64-68

Collini E.,Wong, C.,Y., Wilk., K.E.,Curmi, P.M.G., Brurner, P., Scholes G.,D., 2010, Coherently wired light harvesting in photosynthetic marine algae at ambient temperature, Nature 463, 644-648.

Crespi B.,J., 2001, The evolution of social behaviour in microorganisms., Trends Ecol. Evol., 16, 178–183.

Crick, F.H.C., 1958, , Symp. Soc. Exp. Biol. XII, 139-163.

Crick, F.,1970, Central dogma of molecular Biology, Nature, 227, 5258, 561-563.

Dee N. D., Kessin,E., Ennis, H. L., 2000, Developmental cheating and the evolutionary biology of Dictyostelium and Myxococcus Microbiology, 146, 1505–1512.

Del Giudice E., Doglia S., Milani M., Vitiello G.., 1986, Electromagnetic field and spontaneous symmetry breaking in biological matter., Nucl. Phys. B 275, 185-199.

Del Giudice, E., Preparata, G..1998, A new QED picture of water: understanding a few fascinating phenomena. In, Sassaroli et al. (Eds.), Macroscopic quantum coherence. World Scientific, London, UK, 108-129.

Del Giudice, E., Vitiello, G., 2006, Role of the electromagnetic field in the formation of domains in the process of symmetry-breaking phase transitions., Phys. Rev. A, 74, 022105.

Devaney R. L. 1989, An Introduction to chaotic dynamical systems, Addison-Wesley.

Douhal A.,Kim S,A.H.,Zewall A.H., 1995, Femtosecond molecular dynamics of tautomerization in model base pairs, Nature, 378, 260-263.

Eigen M., Schuster P., 1979, The Hypercycle,, A principle of natural self-organization. Springer, Berlin.

Einstein A., Podolsky B. Rosen N., 1935, Can Quantum-Mechanical Description of Physical Reality be Considered complete?, Phys. Rev., 41, 777.

Eldredge N., 2008, Hierarchies and the sloshing bucket: toward the unification of evolutionary Biology, Outreach, 1:10-15.

Eldredge, N., Gould, S. J.,1972, Punctuated equilibria: an alternative to phyletic gradualism. In: Schopf, J. M. editor, Models in paleo-Biology. San Francisco, W. H. Freeman, 72: 82–115.,

Engel,G.S., Calhoun, T.R., Read, E.L., Ahn, T., Mançal, T., Cheng Yuan-Chung, Blankenship R.E., Fleming,G.R., 2007, Evidence for wavelike energy transfer through quantum coherence in ptotosynthetic systems, Nature, 447, 782-786

Flowers, I., Si I. Li., Stathos,A.,.¤, G.Saxer, E. A. Ostrowski, D. C.,J.E.Strassman, M.D. Purugganan,2010, Variation, sex, and social cooperation: molecular population Genetics of the social amoeba Dictyostelium discoideum, PLoS Genetics, 6,7,1-14

Fox Keller, E., 2005, Revisiting "scale free" networks. Bioessays, 27:1060-1068.

Queller, J. E. Strassmann, M. D., Purugganan, 2010, Variation, Sex, and Social Cooperation: Molecular Population Genetics of the Social Amoeba Dictyostelium Discoideum, PLoS Genetics, 6,1-14.

Fisher, R.A., 1930, The genetical theory of selection, Clarendon, Oxford.

Gaspard P., 2007, Time asymmetry in non equilibrium statistical mechanics, Advances in Chemical Physics, 135, 83-133.

Gerhart,J., Kirschner, M., 1999, Cells, Embryos and Evolution, Blackwell Science, Mass. USA.

Gould S.J., 1998, Full House, New York, Harmony Books.

Gray, H, and Winkler, J.,2003, "Electron tunneling through proteins." Q. Rev. Biophys., 36, 341–372,.

Huxley, J., 1943, Evolution, the modern synthesis, Harper and Brothers Publishers, New York and London,.

Jaeger G., 2009, Entanglement, information, and the interpretation of quantum mechanics Heildelberg: Springer.

Jean R. V., 1994, Phyllotaxis : A systemic study in plant morphogenesis, Cambridge University Press,.

Jeong H., B., Tombor, R. Albert, Z.N. Ottvai, A.L. Barabasi,J., 2000, The large scale organization of metabolic networks, Nature, 407,651.

Karafyllidis,,I.G., 2008, Quantum mechanical model for information transfer from DNA to protein. Biosystems, 93, 191-198 *

Kucharski,R., Maleszka, R., Foret,S., Maleszka,A.,2008, Nutritional control of reproductive status in Honeybees via DNA Methylation, Science, 319: 1827-1830.

Laskar J., 1990,"The chaotic behaviour of the solar system", Icarus, 88, 266-291.

Laskar J.,1994, Large scale chaos in the Solar System, Astron. Astrophys., 287, 9 -12.

Lerner I.M., Population genetics and animal improvement, Cambridge University Press, 1950.

Longo,G., Palamidessi,C., Thierry P.,2010, Some bridging results and challenges in classical, quantum and computational randomness. In Randomness through computation, H. Zenil (ed), World Sci.,.

Longo G., 2011, Interfaces de l'incomplétude, Les Mathématiques, Editions du CNRS, (ongoing translation in English).

Longo G., Montévil M., 2011, From physics to biology by extending criticality and symmetry reakings. To appear in a special issue of Progress in Biophysics and Molecular Biology,.

Longo G., Montévil M.,2012, Anti-entropy as evo/devo order and disorder. *In preparation,*.

Lush J.L., 1945, Animal breeding plans., Iowa D State College Press, Ames USA.

Mehdiabadi,N.,J., Kronforst,M.R., . Queller,D.C, and Strassmann,J.,2008, Phylogeny, reproductive isolation and kin recognition in the social amoeba Dictyostelium Purpureum Evolution 63-2: 542–548.

Mather K.,1949, Biometrical Genetics. Methuen and Co. London,.

Michael, J.C., 1992, Bacterial differentiation within Maraxella bovis colonies growing at the interface of the agar medium with the Petri dish, J.General Microbiology, 138, 2687-2695 *

Monod J.,1970, L'hasard et la nécessité, Seuil, Paris,.

Mukamel R.,Ekstrom A.D., Kaplan, J.,Iacoboni M., Fried, I.,2010, Single neuron responses in humans during execution and observation of actions, Current Biology, 20, 750-756,

Nielsen, MA, and Chuang, IL,2000, Quantum computation and quantum information, Cambridge University Press, Cambridge, U.K.,.

Nilsson-Ehle,,1909, Kreuzunguntersuchungen an Hafer und Weizen. Academic Dissertation, Lund,122pp,.

Noguera, M., Bertran,J:,Sodupe, M.,2004, A quantum chemical study of Cu2+ iteracting with guanine-cytosine base pair . Electrostatic and oxidative effects on intermolecular proton-transfer processes, Journal of Physical Chemistry, 108 :333-341,.

Perez A., Tuckerman, M.E., Hjalmarson. H.P., von Lilienfeld O.A., 2010, Enol tautomers of Watson-Crick base pair models are metastable because of nuclear quantum effects, Journal of the American Chemical Society, 132: 11510-11515

Ostrowski, E.A., Katoh M, Shaulsky,G.,, Queller,D.C., Strassmann, J.E.,2008, Kin discrimination increases with genetic distance in a social amoeba, PLoS Biology, 6: 2377-2382,.

Poincaré H.,1892, Les méthodes nouvelles de la mécanique celeste, Paris.

Reece, St.Y., and Nogera D.G. 2009, Proto-coupled electron transfer in Biology: results from synergistic studies in natural and model systems. Annual Review of Biochemistry, 78: 673-699,.

Ricard J.,2008, Pourquoi le tout est plus que la somme de ses parties, Hermann, Paris.

Rieger T, Neubauer Z, Blahůško A, Cvrčkov,F., Markoš. A.2008,. Bacterial body plans: colony ontogeny in Serratia marcescens. Communicative Integrative Biology, 1:78-87.

Sarovar, M., Ishizaki, A., Fleming, G.R., Whaley K.B.,2010, Quantum entanglement in photosynthetic light-harvesting complexes, Nature-Physics, 6, 462-467.

Schrödinger, E. What Is Life?,1944, The Physical Aspect of the Living Cell, Cambridge University Press, Cambridge, U.K..

Sonnenschein C, Soto A.M..1999, The society of cells: cancer and control of cell proliferation. New York: Springer Verlag.

Soto A.M., Sonnenschein C., Miquel P.A. 2008, On physicalism and Downward Causation in Developmental and Cancer Biology, Acta Biotheor, 56:257–274.

Strassmann J.E., Yong Zhu, Queller, D.C.,2000, Altruism and social cheating in the social amoeba Dictyostelium discoideum Nature 408: 465-466.

Van Driessche, N, Shaw,C., Katoh,M., Morio,T., Sucgang, R., Ibarra,M., Kuwayama,H., SaitoT., Urushihara,H., Maeda,M., Takeuchi,I., Ochiai,H., Eaton,W., Tollett, J., Halter,J., Kuspa,A., Tanaka.Y,.Shaulsky,G.,2002, A transcriptional profile of multicellular development in Dictyostelium discoideum, Development 129, 1543-1552.

Vries de H.1902 Die Mutationstheorie, Veit, Lipsia,.

Waddington, C.H., 1975, The evolution of an evolutionist, Cornell University Press

Watson J.D., Crick, F. H.C.,1953, Molecular structure of nucleic acids: a structure for deoxyribonucleic acid, Nature, 171:737-738.

West SA, Griffin AS, Gardner A, Diggle SP.,2006 , Social evolution theory for microorganisms. Nat Rev Microbiol 4: 597–607.

Winkler, J, Gray, H, Prytkova, T, Kurnikov, I, and Beratan, D.,2005, "Electron transfer through proteins." Bioelectronics, pp 15–33, Wiley-VCH,Weinheim, Germany.

Wimpenny, J.W.T, Parr, J.A.1979, Biochemical Differentiation in Large Colonies of Enterobacter cloacae, Journal of General Microbiology, 114, 487-489.

Wright S., 1932, The roles of mutation, imbreeding, crossbreeding ad selection in evolution. Proc.sixth. Int.Congr. Genet.. 1:356-366

Zak M., From quantum entanglement to mirror neurons, 2007, Chaos, Solitons and Fractals 34, 344–359,.

Zeh, D.W,. Zeh,J.A., and Yoichi Ishida,2009, Transposable elements and an epigenetic basis for punctuated equilibria.31,715-726.